\documentclass{nature}
\usepackage{graphicx}
\usepackage{pdflscape}
\usepackage{hyperref}
\usepackage{threeparttable}
\usepackage{xcolor}
\usepackage{ulem}

\usepackage{caption}
\usepackage[caption = false]{subfig}
\usepackage{amsmath, amssymb}

\definecolor{orcidlogocol}{HTML}{A6CE39}

\makeatletter
\renewcommand{\baselinestretch}{1.2} 

\let\saved@includegraphics\includegraphics
\AtBeginDocument{\let\includegraphics\saved@includegraphics}
\renewenvironment*{figure}{\@float{figure}}{\end@float}
\makeatother

\newcommand{\apjl}{Astrophys. J.}
\newcommand{\apjs}{Astrophys. J.}
\newcommand{\aap}{Astron. Astrophys.}

\def\be{\begin{eqnarray}}
\def\ee{\end{eqnarray}}

\makeatletter
\renewcommand{\baselinestretch}{1.2} 

\let\saved@includegraphics\includegraphics
\AtBeginDocument{\let\includegraphics\saved@includegraphics}
\renewenvironment*{figure}{\@float{figure}}{\end@float}
\makeatother

\makeatletter
\def\@fnsymbol#1{\ensuremath{\ifcase#1\or \dagger\or \ddagger\or
 \mathsection\or \mathparagraph\or \|\or **\or \dagger\dagger
 \or \ddagger\ddagger \else\@ctrerr\fi}}
\makeatother
 
\newcommand{\FIG}[1] { Figure~\ref{#1}}
\newcommand{\TAB}[1] {Table~\ref{#1}}
\newcommand{\EXTTAB}[1] {Extended Data Table~\ref{#1}}
\newcommand{\EXTFIG}[1] {Extended Data Figure~\ref{#1}}

\usepackage[utf8]{inputenc}
\usepackage{hyperref}
\usepackage{graphicx}
\usepackage{longtable}
\usepackage{supertabular}
\usepackage{float} 

\title{No pulsed radio emission during a bursting phase of a Galactic magnetar}

\author{L. Lin*$^{1}$, C. F. Zhang*$^{2,3}$, P. Wang*$^{3}$, H. Gao$^{1}$, X. Guan$^{3}$, J. L. Han$^{3,5}$, J. C. Jiang$^{2,3}$, P. Jiang$^{3}$, K. J. Lee$^{4,3}$\thanks{E-mail: kjlee@pku.edu.cn,\href{https://orcid.org/0000-0002-1435-0883}{\textcolor{orcidlogocol}{} \hspace{2mm} orcid.org/0000-0002-1435-0883}}, D. Li$^{3,5}$\thanks{Email:dili@nao.cas.cn,\href{https://orcid.org/0000-0003-3010-7661}{{\textcolor{orcidlogocol}{} \hspace{2mm} orcid.org/0000-0003-3010-7661}  }}, Y. P. Men$^{2,3}$, C. C. Miao$^3$, C. H. Niu$^3$, J. R. Niu$^3$, C. Sun$^3$, B. J. Wang$^{2,3}$, Z. L. Wang$^3$, H. Xu$^{2,3}$, J. L. Xu$^3$, J. W. Xu$^{2,3}$, Y. H. Yang$^6$, Y. P. Yang$^7$, W. Yu$^8$, B. Zhang$^9$\thanks{Email: zhang@physics.unlv.edu,\href{https://orcid.org/0000-0002-9725-2524}{{\textcolor{orcidlogocol}{} \hspace{2mm}orcid.org/0000-0002-9725-2524} }}, B.-B. Zhang$^{6,10,9}$, D. J. Zhou$^{5,3}$, W. W. Zhu$^3$, A. J. Castro-Tirado$^{11,12}$, Z. G. Dai$^{6,10}$, M. Y. Ge$^{13}$, Y. D. Hu$^{11,14}$, C. K. Li$^{13}$, Y. Li$^{4,15}$, Z. Li$^1$, E. W. Liang$^{16}$, S. M. Jia$^{13}$, R. Querel$^{17}$, L. Shao$^{18}$, F. Y. Wang$^{6,10}$, X. G. Wang$^{16}$, X. F. Wu$^{15}$, S. L. Xiong$^{13}$, R. X. Xu$^{2,4}$, Y.-S. Yang$^6$, G. Q. Zhang$^6$, S. N. Zhang$^{13,3,5}$, T. C. Zheng$^{16}$, J.-H. Zou$^{18}$}

\begin{document}
\maketitle
\begin{affiliations}
 \item Department of Astronomy, Beijing Normal University, Beijing 100875 , P.R.China
 \item Department of Astronomy, Peking University, Beijing 100871, P.R.China
 \item National Astronomical Observatories, Chinese Academy of Sciences, Beijing 100101, China
 \item Kavli institute for astronomy and astrophysics, Peking University, Beijing. 100871, P.R.China
 \item University of Chinese Academy of Sciences, Chinese Academy of Sciences, Beijing 100049, China 
 \item School of Astronomy and Space Science, Nanjing University, Nanjing 210093, China
 \item South-Western Institute for Astronomy Research, Yunnan University, Kunming 650500, Yunnan, China
 \item Shanghai Astronomical observatory, Chinese Academy of Science, Shanghai 200030, China
 \item Department of Physics and Astronomy, University of Nevada, Las Vegas, NV 89154, USA
 \item Key Laboratory of Modern Astronomy and Astrophysics (Nanjing University), Ministry of Education, China
 \item Instituto de Astrof\'isica de Andaluc\'ia (IAA-CSIC), Glorieta de la Astronom\'ia s/n, E-18008, Granada, Spain
 \item Departamento de Ingenier\'ia de Sistemas y Autom\'atica, Escuela de Ingenier\'ias, Universidad de M\'alaga, Dr. Pedro Ortiz Ramos, 29071 M\'alaga, Spain
 \item Key Laboratory of Particle Astrophysics, Institute of High Energy Physics, Chinese Academy of Sciences, Beijing 100049, China
 \item Universidad de Granada, Facultad de Ciencias Campus Fuentenueva S/N CP 18071 Granada, Spain
 \item Purple Mountain Observatory, Chinese Academy of Sciences, Nanjing 210023, China
 \item Guangxi Key Laboratory for Relativistic Astrophysics, School of Physical Science and Technology, Guangxi University, Nanning 530004 China;
 \item National Institute of Water and Atmospheric Research (NIWA), 
Lauder, New Zealand
 \item College of Physics, Hebei Normal University, Shijiazhuang 050024, China
 \end{affiliations}
\noindent{*These authors contributed equally to this work}

\bigskip

\begin{abstract} 
\bf Fast radio bursts (FRBs) are mysterious millisecond-duration radio transients of unknown origin observed at extragalactic distances\cite{lorimer07,petroff19,cordes19}. It has been long speculated that magnetars are the engine powering repeating bursts from FRB sources\cite{popov10,kulkarni14,murase16,katz16,metzger17,beloborodov17,kumar17,yangzhang18,wadiasingh20,cheng20}, but no convincing evidence has been collected so far\cite{sun19}. Recently, the Galactic magnetar SGR J1935+2154 entered an active phase by emitting intense soft $\gamma$-ray bursts\cite{SGR-GBM}. One FRB-like event with two peaks (FRB 200428) and a luminosity slightly lower than the faintest extragalactic FRBs was detected from the source\cite{SGR-CHIME,SGR-STARE2}, in association with a soft $\gamma$-ray / hard X-ray flare\cite{SGR-HXMT,2020ApJ...898L..29M,SGR-AGILE,SGR-Konus}. Here we report an eight-hour targeted radio observational campaign comprising four sessions and assisted by multi-wavelength (optical and hard X-rays) data. During the third session, 29 soft $\gamma$-ray repeater (SGR) bursts were detected in $\gamma$-ray energies. Throughout the observing period, we detected no single dispersed pulsed emission coincident with the arrivals of SGR bursts, but unfortunately we were not observing when the FRB was detected. The non-detection places a fluence upper limit that is eight orders of magnitude lower than the fluence of FRB 200428. Our results suggest that FRB -- SGR burst associations are rare. FRBs may be highly relativistic and geometrically beamed, or FRB-like events associated with SGR bursts may have narrow spectra and characteristic frequencies outside the observed band. It is also possible that the physical conditions required to achieve coherent radiation in SGR bursts are difficult to satisfy, and that only under extreme conditions could an FRB be associated with an SGR burst. 
\end{abstract}

\begin{figure}[H]
 \centering
 \includegraphics[width=6.0in]{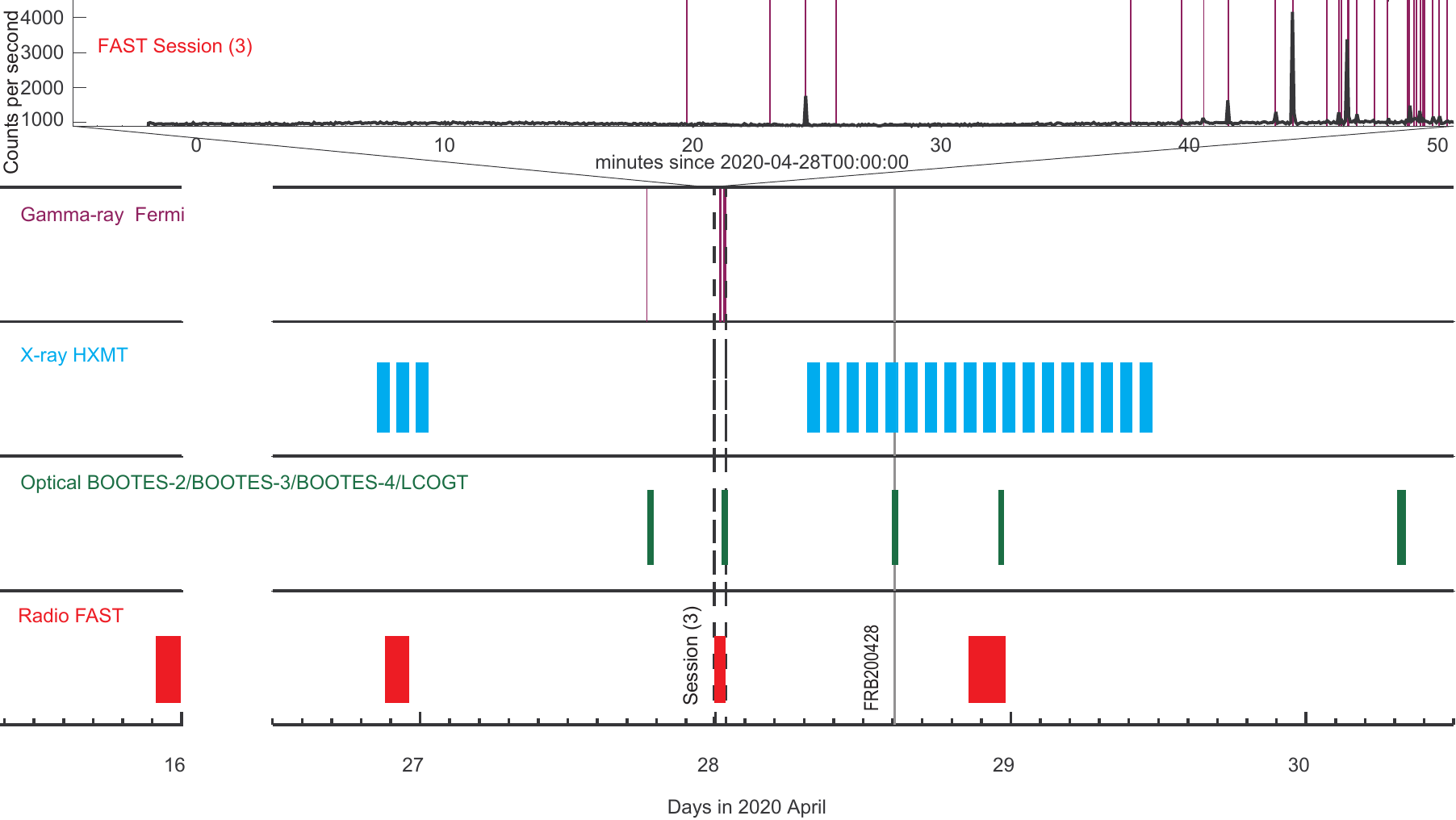}
 \caption{{\bf Timeline of our multi-wavelength observational campaign of SGR J1935+2154.} Each observation is labelled with a different colour and colour blocks indicate the observing epochs. In the $\gamma$-ray band, the colour blocks indicate the observing epochs. The zoomed-in timeline shows the detections of 29 SGR bursts by GBM in the third FAST session. The epoch of FRB 200428 is marked as a grey line, which coincides with one of observations of BOOTES. The second and third optical observations are artificially thickened for clarity.}
 \label{fig:timeline}
\end{figure}

We have been closely monitoring SGR 1935+2154 with FAST\cite{nan11} to test whether a magnetar can create FRBs or FRB-like events during its active phase. We observed the target for 8 h in the following four sessions: (1) 15 April 2020 21:54:00 to 23:54:00 UTC (coordinated universal time), (2) 26 April 2020 21:06:55 to 23:06:55 UTC, (3) 27 April 2020 23:55:00 to 28 April 2020 00:50:37 UTC; and (4) 28 April 2020 20:35:00 to 23:35:00 UTC. These observing windows are uneven because they are limited by the visibility of the source and the availability of observing time with FAST. We used the FAST central beam of the L-band receiver with a usable 460 MHz band centered around 1.25 GHz. The system temperature was 20-25 K. During the FAST observing period, we also coordinated a multi-wavelength observational campaign in the hard X-ray band with {\it Insight}-HXMT and in the optical band with the BOOTES telescopes in China, Spain and New Zealand, as well as the LCOGT 1-m telescope in USA (Fig.\ref{fig:timeline}).

FRB 200428\cite{SGR-CHIME,SGR-STARE2} occurred between sessions (3) and (4), so the signal was not caught by FAST. On the other hand, during the 1-h observing period in session (3), SGR 1935+2154 became very active, emitting 29 bursts in about 30 minutes (Methods). The temporal and spectral properties of these events are similar to those of standard magnetar short bursts observed with Fermi/GBM\cite{collazzi15,lin2020}. The fluence of these bursts in $8-200$~keV is in the range of $1.8\times10^{-8}\sim6.7\times10^{-6}~{\rm erg}~{\rm cm}^{-2}$. Some of these bursts have fluences comparable to or higher than $(6.8\pm0.1)\times10^{-7}~{\rm erg}~{\rm cm}^{-2}$, the 1-250 keV fluence of the burst associated with FRB 200428\cite{SGR-HXMT} . Considering the huge fluence ($>1.5$ MJy~ms\cite{SGR-STARE2}) of FRB 200428, if associations of FRBs with SGR bursts were ubiquitous, one would expect the detection of at least 29 FRBs in session (3). 

\begin{figure}[H]
 \centering
 \includegraphics[width=6.0in]{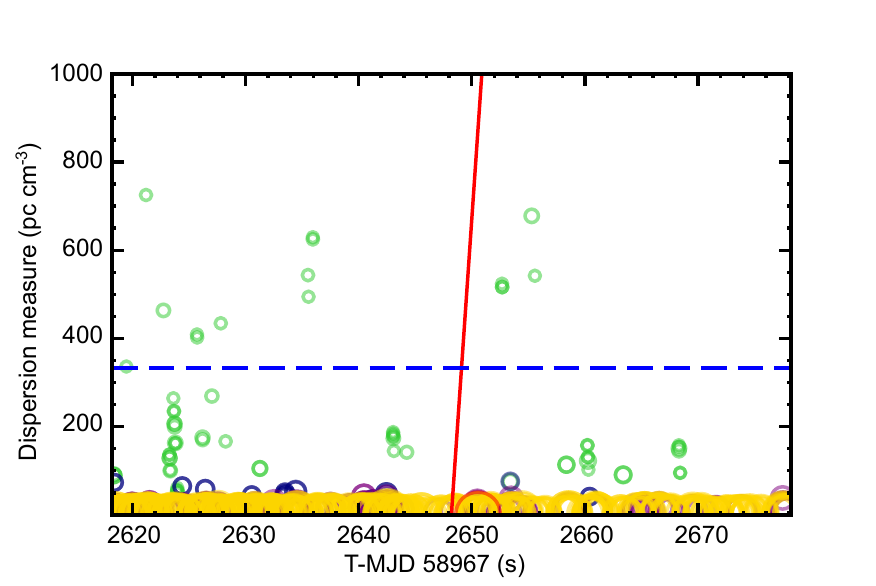}
 \caption{{\bf Non-detection of radio burst within $\pm 30$s of the GBM burst 10.} The horizontal and vertical axes show the observational time and the DM, respectively. The size and colour of the markers indicate the flux density of the signal; that is, green: ${\rm flux}<5$mJy, blue: $5 {\rm mJy} <{\rm flux}\le10 {\rm mJy}$, purple: $10 {\rm mJy} <{\rm flux}\le20 {\rm mJy}$, yellow: $20 {\rm mJy} <{\rm flux}\le40 {\rm mJy}$, red: $40 {\rm mJy} <{\rm flux}$. Throughout the observations, no signal with apparent flux density above 80 mJy was detected. The slanted red line is the expected arrival time of a putative FRB associated with the SGR burst, and the horizontal dashed blue line indicates DM = 333 $\rm pc \ cm^{-3}$. The plot shows the DM range from 0 -- 1,000 $\rm pc \ cm^{-3}$ with a searching range of 0 -- 5000 $\rm pc \ cm^{-3}$. MJD, modified Julian date.
 \label{fig:radio-3SGRburst}}
\end{figure}

We performed dedicated searches for FRB-like events in all four sessions, paying special attention to session (3), when 29 SGR bursts were emitted. Because a megajansky-level radio burst would saturate FAST, we searched for both dispersion signals and instrumental saturation signals. Five types of searching strategies were applied: blind search, limited dispersion measure (DM) search, saturation search, windowed search, as well as ephemeris folding (Methods). No single burst signal consistent with the SGR 1935+2154 origin was detected down to the FAST sensitivity limit. \FIG{fig:radio-3SGRburst} shows the candidate single pulses detected as a function of time and DM for one example Fermi/GBM burst (No. 10). The red solid line tracks the expected FRB arrival time as a function of DM at 1.25 GHz. The horizontal blue dashed line denotes DM $\simeq 333 \ {\rm pc \ cm^{-3}}$, measured from the FRB-like events detected from the source\cite{SGR-CHIME}. One can see that no signal was detected at the desired time and DM. We checked all the pulse candidates one by one and identified them as narrow-band radio frequency interferences (RFIs). The same is true for all other 28 Fermi/GBM bursts detected in the same observing session (Methods and \EXTFIG{fig:all-radio-SGRburst}). The non-detection of bursts from SGR 1935+2154 by FAST sets stringent upper limits on the fluxes of pulsed radio emission as low as several millijanskys (\EXTFIG{fig:limits}) and on the fluences down to the level of 10-40 mJy ms (\EXTFIG{fig:limits}).

The lack of any FRBs in association with any of the 29 SGR bursts poses important constraints on the physical mechanism producing observable FRBs from magnetars. We consider the following three possibilities, noting that more than one of these may be responsible for the missing FRBs from most SGRs.

The first possibility is that all SGR bursts may be associated with FRBs, but the FRB jets are much more collimated than high-energy emission so that most of them missed Earth. 
Let us assume that each FRB has a conical structure with a half-opening angle defined as max $(1/\Gamma, \theta_j)$, where $\Gamma$ and $\theta_j$ are the Lorentz factor and geometric opening angle of the FRB jets, respectively. The fact that at most 1/30 of the SGR bursts have a detectable FRB down to the FAST flux sensitivity (considering the detection of FRB 200428\cite{SGR-CHIME,SGR-STARE2}, which is associated with one SGR burst\cite{SGR-HXMT,2020ApJ...898L..29M,SGR-AGILE,SGR-Konus}, and that no other FRB associated with other SGR bursts that are not in the FAST observing windows has been detected) suggests that the solid angle of the FRB beam must be at most 1/30 of that of SGR bursts. Assuming that the SGR emission is isotropic, one reaches the most conservative constraints of $\Gamma \geq 59$ and $\theta \leq 0.37$ rad, which should be more stringent if SGR bursts themselves are beamed (Methods). The requirement of $\Gamma \gg 1$ is consistent with the suggestion that FRB emitters must be highly relativistic according to other theoretical arguments\cite{murase16,lu18}. If beaming is the cause of the non-detection of FRBs from most SGR bursts, the true energies and luminosities of FRBs should be corrected by the small beaming factor $f_{\rm b,FRB}$ and are much lower than the isotropic values. For FRB 200428, the energy is $<10^{34}$ erg and the luminosity is $< 3\times10^{36} \ {\rm erg \ s^{-1}}$ (Methods).

The second possibility is that all SGR bursts are accompanied by low-frequency bursts, but the peak frequencies of these bursts may have a range of distribution. This scenario may apply to the FRB models invoking relativistic shocks and contrived conditions to produce synchrotron maser emission\cite{lyubarsky14,waxman17,metzger19,beloborodov20}. In order to have at most 1/30 of SGRs producing FRBs observable by FAST, very contrived conditions are necessary. The required spectra of low-frequency bursts must be extremely narrow and the distribution of the peak frequencies of these bursts must be far from the FAST band. If the discrepancy between CHIME\cite{SGR-CHIME} and STARE2\cite{SGR-STARE2} for the fluence of FRB 200428 is caused by the intrinsic narrow spectrum of the FRB, then the distribution peak of the putative low-frequency bursts associated with other 29 SGR bursts should be above 30 GHz (Methods).

The final possibility is that the observed rarity of FRBs from SGR bursts is intrinsic. The extremely high brightness temperatures of FRBs require that the radiation mechanism must be coherent\cite{lyubarsky14,kumar17,lu18,yangzhang18,waxman17,metzger19,beloborodov20}. It is possible that the fragile coherent condition may not always be satisfied in SGR bursts. For this possibility, one would expect that the SGR burst associated with FRB 200428 has some special features that are uncommon in most SGR bursts. Tentative evidence along this line has been collected\cite{SGR-HXMT}.

The non-detection of FRBs from 29 SGR bursts is consistent with the known rates of SGR bursts and cosmological FRBs. If all SGR bursts similar to the one associated with FRB 200428 generate FRBs with luminosities similar to that of FRB 200428, the cosmological FRB rate would be about two orders of magnitude higher than the currently observed value (Methods). This discrepancy is fully consistent with our observation, which shows that at most 1/30 SGR bursts produce observable FRBs.

Our multi-wavelength observations set upper limits in their respective observing windows (Methods). Most of these upper limits are not constraining, but one $Z$-equivalent 17.9 magnitude upper limit in the 60-s exposure during the prompt epoch of FRB 200428 set by BOOTES-3 can pose some interesting constraints on the models for the so-called ``fast optical bursts'' (FOBs) associated with FRBs\cite{yang19}. In particular, this upper limit can rule out certain parameter space of some FOB models invoking the inverse Compton scattering origin of optical emission (Methods).

\bigskip
\bigskip
\bigskip
\bibliography{FRB}

\begin{addendum}
\item This work
 is supported by the Natural Science Foundation of China (grants 11988101, 11673002, 11703002, 11543004, 11722324, 11690024, 11633001, 11920101003, 11833003, 11722324, 11633001, 11690024, 11573014, 11725314, 11690024, 11743002, 11873067, 11533003, 11673006, U1938201, 11725313,  11721303, 1172130, U15311243, U1831207, U1838201, U1838202, U1838113, U1938109), National Key Research and Development Programs of China (NO. 2018YFA0404204, 2017YFA0402600), the Program for Innovative Talents and Entrepreneur in Jiangsu and the KIAA-CAS Fellowship and the China Postdoctoral Science Foundation (No. 2018M631242). 2016YFA0400800, 2018YFA0400802, E01S11BQ10, XDB23010200,XDB2304040, QYZDY-SSW-SLH008, the International Partnership Program of Chinese Academy of Sciences grant No.\ 114A11KYSB20160008, Cultivation Project for FAST Scientific Payoff and Research Achievement of CAMS-CAS, the Max-Planck Partner Group, the Spanish Science Ministry ``Centro de 
Excelencia Severo Ochoa'' Program under grant SEV-2017-0709, and the Junta de Andaluc\'ia (Project 
P07-TIC-03094) and support from the Spanish Ministry Projects 
AYA2012-39727-C03-01, AYA2015-71718R and PID2019-109974RB-I00. We thank E. Fern\'andez-Garc\'ia (IAA-CSIC), I. M. Carrasco-Garc\'ia and C. P\'erez del Pulgar (UMA) and the rest of the BOOTES Team members for making the reported BOOTES Network observations possible.
This work made use of data from the FAST. FAST is a Chinese
 national mega-science facility, built and operated by the National
 Astronomical Observatories, Chinese Academy of Sciences. We also acknowledge the use of public data from the Fermi Science Support Center (FSSC).
 \item[Author Contributions] 
LL, BZ, and DL launched the FAST observational campaign on SGR J1935+2154; CFZ and PW systematically processed the FAST data independently and cross compared the results; KJL and DL coordinated FAST data analysis campaign; JLH, YPM, CCM, CHN, JRN, BJW, HX, JLX, WY, DJZ, and WWZ participated in the FAST data analysis. XG, PJ, CS, and ZLW coordinated FAST observations; BBZ, LL, YHY, LS, YSY, JHZ, FYW, and GQZ processed {\it Fermi/}GBM data; LL, SNZ, MYG, SMJ, CKL, and SLX performed joint {\it Insight}-HXMT observations with FAST and processed the data; BBZ, AJC-T, YDH, and RQ carried out the BOOTES optical observations; XGW, EWL, and TCZ carried out the LCOGT optical observations; BZ coordinated the science team; HG, YPY, ZGD, YL, ZL, FYW, XFW, RXX contributed to theoretical investigations of the physical implications of the observational results. BZ, KJL, BBZ, YPY, HG, LL, CFZ, YL, and JLH contributed to the writing of the paper.

 \item[Competing Interests] The authors declare that they have no
competing financial interests.

\item[Correspondence] Correspondence and requests for materials
should be addressed to K. J. Lee (kjlee@pku.edu.cn), D. Li (dili@naoc.ac.cn), and B. Zhang (zhang@physics.unlv.edu).

 \item[Data Availability Statement]
 Processed data are presented in the tables and figures in the paper. Source data are available upon reasonable requests to the corresponding authors. The {\it Fermi}/GBM data are publicly available at  \url{https://heasarc.gsfc.nasa.gov/FTP/fermi/data/}.
 
\end{addendum}

\clearpage

\newpage

\section*{Methods}

\subsection{Multi-wavelength campaign}

The information of our multi-wavelength observational campaign is presented in \EXTTAB{tab:timeline}. \FIG{fig:timeline} shows how the multi-wavelength observations overlap in time.

\subsection{FAST observations}
We performed four sessions of FAST observations as listed in \TAB{tab:timeline}: 
The centre frequency is 1.25 GHz, spanning from 1.0 GHz to 1.5 GHz, including a 20-MHz band edge on each side. The average system temperature was 25 K.

We searched for radio bursts with either a dispersion signature or instrumental saturation to all FAST data collected during the observational campaign.
The search for dispersed bursts was carried out using the software package BEAR\cite{MLC19}. Five types of searches had been performed: blind search, dedicated search, saturation search,  windowed search, and ephemerous folding.

{\em Blind search:} We searched the DM range 1 -- 5,000 $\rm pc\, cm^{-3}$. We used the following scheme to de-disperse the data to save computational resource: DM steps of 0.3, 0.5, 1.0, and 3.0 $\rm pc\, cm^{-3}$ were used for the DM ranges $0-1,000$, $1,000-1,835$, $1,835-3,656$, and $3,656-5,000$ $\rm pc\, cm^{-3}$, respectively. We used 14 box-car-filter-width grids uniformly distributed in logarithmic space from 0.2 ms to 30 ms. A zero-DM matched filter\cite{MLC19} was applied to mitigate RFI in the blind search. All the candidate plots generated were then visually inspected. We found that most of the candidates were narrow-band RFIs, and there was no burst with dispersive signature with a signal-to-noise ratio of $\rm S/N \ge 8$. The DM-time plots for the 29 bursts seen in the {\it Fermi}/GBM (Gamma-Ray Burst Monitor) daily time-tagged event (TTE) data are presented in \EXTFIG{fig:all-radio-SGRburst}.

{\em Dedicated search:} For the DM range from 200 -- 600 $\rm pc\, cm^{-3}$, we performed a 
dedicated search. The DM step was refined to $0.3~\rm pc\,cm^{-3}$. In the dedicated search, we shut down the RFI mitigator to make sure we did not discard signals. We did not detect any wide-band dispersive bursts with $\rm S/N \ge 8$.

{\em Saturation search:} We understand that if the radio flux is as high as kilojansky -- megajansky\cite{SGR-CHIME,SGR-STARE2}, FAST would be saturated.
We therefore also searched for saturation signals in the data. We looked for the epoch in which 50\% of channels satisfy one of the following conditions: 1) the channel is saturated (255 value in 8-bit channels), 2) the channel is zero-valued, or 3) the root mean square of the bandpass is less than 1. Because the DM for SGR 1935+2154 is about\cite{SGR-CHIME,SGR-STARE2} $\rm 333\, pc\, cm^{-3}$, we expect that the timescale for saturation is $\sim$800 ms across the FAST band. Our code captured several short timescale wide-band saturation signals. However, we did not detect any saturation lasting longer than \textgreater\, 0.5s. We exclude any saturation associated with SGR 1935+2154.

{\em Windowed search:} 
The blind search in the {\it Fermi}/GBM daily TTE data during the FAST observing time revealed 29 short SGR-like bursts on 28 April 2020 from 00:19 to 00:50 UTC (\EXTTAB{tab:fermitab}). 
For the FAST data within a time window of $\pm 1$ min of these {\it Fermi}/GBM events, we performed a windowed search, in which we visually inspected the plots of 
the dynamic spectra, the DM-0 time series, and the de-dispersed time series at 333 $\rm pc\, cm^{-3}$. No burst or saturation signal consistent with an SGR 1935+2154 origin was detected. 

This non-detection places stringent upper limits on the pulsed radio emission from SGR 1935+2154. 
Using the telescope parameters\cite{Jiang20RAA}, for a 5-ms pulse the estimated flux and fluence upper limits are 4.5 mJy and 22 mJy ms, respectively (see \EXTFIG{fig:limits}).

For the third session, we also folded the data using the ephemeris information of SGR 1935+2154\cite{israel16} and the DM value reported\cite{SGR-CHIME,SGR-STARE2}. No obvious signal was seen.

\subsection{Fermi Observations of SGR J1935+2154}

We performed a blind search with the Bayesian Blocks method\cite{lin2020} using the continuous time-tagged event data of GBM. We found 29 bursts from SGR $1935+2154$ during the third session of FAST observations. All these events were detected on 28 April 2020 in about 30 min, from 00:19:44.192 to 00:50:21.969. Both the flux and the fluence are in the range $8-200$~keV. The distribution of the burst durations (with mean value $\left < T_{\rm 90} \right >\sim0.22$~s, where $T_{90}$ is the burst duration defined by the time window in which 90\% of the burst fluence is collected) and the cumulative distribution of the fluence is presented in \EXTFIG{fig:t90_dis}. A cut-off power-law function
 \be
 N(>S) = A\left[ \left( \frac{S}{10^{-7}} \right)^{\alpha} -
 \left(\frac{S_{\rm max}}{10^{-7}}\right)^{\alpha}\right]
 \ee
is used to fit the cumulative distribution of fluence, where $S$ and $S_{\rm max}$ are the fluence and maximum fluence, respectively. Using the Markov chain Monte Carlo method, the best-fit parameters are $ \alpha = 0.38^{+0.06}_{-0.07} $
and $ S_{\rm max} = 1.67^{+1.06}_{-0.76} \times 10^{-5} $ erg cm$ ^{-2} $.

\subsection{Hard X-ray observations}

During the FAST observing sessions (2) and (4), the {\it Insight}-HXMT (Hard X-ray Modulation Telescope) X-ray satellite\cite{HXMT} observed SGR 1935+2154 simultaneously with  its three collimated telescopes covering the 1-250 keV energy band. No significant (above $3\sigma$) detection of any burst was made in our offline data analysis. Assuming a cut-off power-law spectral model with the same parameters (neutral hydrogen column density  $2.6\times10^{22}~{\rm cm}^{-2}$; photon index 1.44; and cut-off energy 69.8~keV) as those used for the hard-X-ray burst detected by {\it Insight}-HXMT in association with FRB 200428\cite{SGR-HXMT} 
from SGR J1935+2124, we obtain the $3\sigma$ upper limit on its fluence as $F_i=A_{i} 10^{-9} T^{-1/2}~{\rm erg~cm}^{-2}$, where $T$ is the assumed burst duration in seconds, $i=1,2,3$ represents the three telescopes, namely, the low-energy X-ray telescope (LE; 1-10 keV), the medium-energy X-ray telescope (ME; 10-30 keV) and the high-energy X-ray telescope (HE; 27- 250 keV) -- and $A_{1}=2.7$, $A_{2}=3.5$ and $A_{3}=4.5$. 
With a duration $T=0.5$ s, similar to the X-ray burst associated with FRB 200428, the upper limits are $3.8 \times 10^{-9}~{\rm erg~cm}^{-2}$, $4.9 \times 10^{-9}~{\rm erg~cm}^{-2}$ and $6.4 \times 10^{-9}~{\rm erg~cm}^{-2}$ for LE, ME and HE, respectively. 

From 15 to 29 April (not included) a period covering all four FAST sessions, about 300 X-ray bursts were observed with several X- or $\gamma-$ray telescopes -- {\it Fermi}/GBM\cite{2020GCN.27659....1F} and {\it Insight}-HXMT\cite{2020GCN.28027....1L}, other detectors include the Burst Alert Telescope\cite{2020ATel13675....1P,2020ATel13758....1T} and the X-Ray Telescope\cite{2020arXiv200600215B} onboard the Neil Gehrels Swift Observatory,  NICER\cite{2020arXiv200611358Y},  AGILE\cite{SGR-AGILE}, Integral\cite{2020ApJ...898L..29M}, Konus-Wind\cite{SGR-Konus}, AstroSat\cite{2020GCN.27664....1M} and the CALET Gamma-ray Burst Monitor\cite{2020GCN.27623....1C}. This number does not include the burst forest containing two shortly separate time intervals of $\sim$3~s and $\sim$15~s long on 27 April, when a series of bursts arrived together and the count rate never returned to the background level\cite{2020ATel13675....1P}. 

\subsection{Optical observations}

We used the BOOTES (Burst Observer and Optical Transient Exploring System; \url{http://bootes.iaa.es}\cite{BOOTES}) robotic telescopes and the 1-m telescope of LCOGT (Las Cumbres Observatory Global Telescope) at McDonald Observatory to monitor SGR 1935+2154 during our FAST monitoring campaign. 

The three BOOTES telescopes (0.6m BOOTES-4/MET robotic telescope at Lijiang Astronomical Observatory, China, 60-cm BOOTES-2/TELMA robotic telescope at IHSM La Mayora, UMA-CSIC, in Algarrobo Costa, Spain, and BOOTES-3 at NIWA Lauder, Otago, New Zealand) reacted to various SGR alerts, and the performed automatic observations of SGR 1935+2154 around the epochs of the FAST monitoring campaign. These observations led to various $3\sigma$ limiting magnitudes by making use of the nearby stars in the USNO-B1.0 and Pan-STARRS catalogues. The results are shown in \EXTTAB{tab:timeline}. Interestingly, a 17.9-mag upper limit was placed with a 60-second exposure time during the epoch in which FRB 200428 was emitted. 
The 1-m telescope of LCOGT at McDonald Observatory took images in the $R$ filter with an 8$\times$300-s exposure, on 30 April at 07:21:48 UT, and an upper limit 21.1 magnitude was derived from the co-added image 
(\EXTTAB{tab:timeline}).

\subsection{Probability for FAST-band FRB-SGR associations}

The fact that 29 SGR bursts monitored by FAST did not show any associated radio bursts down to the FAST sensitivity level, along with the fact that FRB 200428\cite{SGR-CHIME,SGR-STARE2} was associated with one SGR flare\cite{SGR-HXMT,2020ApJ...898L..29M,SGR-AGILE,SGR-Konus}, can define the baseline probability of detecting an FRB from an SGR burst to be
\be
P_0=1/30.
\ee 
The true probability is probably $P \leq P_0$, because no other FRB has been reported to be associated with any other SGR bursts from SGR 1935+2154. If an FRB-like event as bright as FRB 200428 occurred in association with any of other $\sim$ 300 X-ray bursts, it would probably have been caught by wide-field radio telescopes such as CHIME (Canadian Hydrogen Intensity Mapping Experiment) or STARE2 (Survey for Transient Astronomical Radio Emission 2). Furthermore, during session (4) of the FAST observations, a few dozen X-ray bursts were detected by NICER\cite{2020arXiv200611358Y}. Because the data of these bursts are not publicly available, we cannot perform a similar analysis to that carried out for the 29 GBM bursts reported here. However, the non-detection of FRBs during session (4) would make $P$ much smaller than $P_0$. In the following, we take $P_0$ as a very conservative probability and discuss the physical implications.

\subsection{Constraints on model parameters: Beaming}

The flux contrast of the 29 non-detections should be 
\be
\frac{F_{\nu,{\rm FRB}}}{F_{\nu,{\rm FAST}}}=\frac{f_{\nu,{\rm FRB}}}{f_{\nu,{\rm FAST}}}\gtrsim\eta\equiv10^8,
\ee
where $f_{\nu,{\rm FRB}}\gtrsim1.5~{\rm MJy~ms}$ is the fluence of FRB 200428, $f_{\nu,{\rm FAST}}\sim22~{\rm mJy~ms}$ is the FAST fluence limit, and $F_{\nu,{\rm FRB}}$ and $F_{\nu,{\rm FAST}}$ are the corresponding fluxes.

We examine the model constraints assuming that this small fraction is caused by the narrow beaming of FRB jets. We consider 
an FRB jet with a geometric beaming angle of $\theta_j$ and a bulk Lorentz factor of $\Gamma$. The Doppler factor at a viewing direction $\theta$ is given by
\be
D(\theta)
=\left\{
\begin{aligned}
&2\Gamma,&~{\rm for}~\theta<\theta_j,&\\
&1/\Gamma[1-\beta\cos(\theta-\theta_j)],&~{\rm for}~\theta>\theta_j. &
\end{aligned}
\right.
\ee
For simplicity, we assume that the intrinsic spectrum of the FRB is flat, which means that the radio flux is independent of frequency (the case of a narrow spectrum is discussed below). Further we assume that FRB 200428 has $\theta<\theta_j$ and the putative SGR-related FRBs not detected by FAST have $\theta>\theta_j$. We then have
\be
\frac{F_\nu(\theta<\theta_j)}{F_\nu(\theta>\theta_j)}\simeq\left(\frac{D(\theta<\theta_j)}{D(\theta>\theta_j)}\right)^3=\left[2\Gamma^2(1-\beta\cos(\theta-\theta_j))\right]^3 \gtrsim\eta
\ee
or $2\Gamma^2(1-\beta\cos(\theta-\theta_j))\gtrsim\eta^{1/3}$. Assuming that $|\theta-\theta_j|\ll1$, one has $1-\beta\cos(\theta-\theta_j)=1-\beta+\beta(\theta-\theta_j)^2/2=1/2\Gamma^2+(\theta-\theta_j)^2/2$, so that $(\theta-\theta_j)^2\Gamma^2\gtrsim\eta^{1/3}$. One may define a characteristic viewing angle 
$\theta_c=\theta_j+\eta^{1/6}/\Gamma$. The above condition is satisfied when $\theta \gtrsim \theta_c$. If we assume that the SGR burst emission is isotropic, the probability for FRB/SGR associations should satisfy
\be
P_0\geq P\simeq\frac{1}{4\pi}\left(2\pi\int_0^{\theta_j+\eta^{1/6}/\Gamma}\sin\theta d\theta\right)=\frac{1}{2}\left[1-\cos\left(\theta_j+\frac{\eta^{1/6}}{\Gamma}\right)\right]\label{beamingp0},
\ee
which gives
\be
\theta_j \leq \arccos\left(1-2P_0\right)-\frac{\eta^{1/6}}{\Gamma}\label{thetagamma}.
\ee
In the left panel of \EXTFIG{fig:beaming}, we plot the constraint from Eq.(\ref{thetagamma}). For $P_0=1/30$, a small beaming angle of $\theta_j\sim0$ means that $\Gamma\gtrsim \eta^{1/6}/\arccos(1-2P_0)\simeq59$, and a large Lorentz factor of $\Gamma\gg\eta^{1/6}$ means that $\theta\lesssim\arccos(1-2P_0)\simeq0.37~{\rm rad}$. In \EXTFIG{fig:beaming}, we plot the function of $P(\theta_j,\Gamma)$ given by Eq.(\ref{beamingp0}). This constraint is very conservative. If the SGR bursts are not isotropic (with a beaming factor $f_{\rm b,SGR} < 1$, then the constrained FRB jet angle would be smaller by the same factor. 

This beaming interpretation, if true, would greatly reduce the required energetics of FRBs. The distance of SGR 1935+2154 is uncertain, ranging from $\sim 6.6$ kpc\cite{zhou20} to $\sim 12.5$ kpc\cite{kothes18}. We adopt the larger value to derive the most conservative upper limits on the true energetics of FRB 200428. According to the STARE2 observation, the isotropic energy of FRB 200428 is $E_{\rm iso,FRB}=4\pi d^2\nu f_{\nu,{\rm FRB}}\sim4\times10^{35}~{\rm erg}$ with $\nu\simeq1.4~{\rm GHz}$ and $d\simeq 12.5~{\rm kpc}$. If the FRB 200428 as detected by STARE2  has the same duration as that  detected by CHIME, that is, $\Delta t\sim5~{\rm ms}$, then its isotropic luminosity is $L_{\rm iso,FRB}\sim8\times10^{37}~{\rm erg~s^{-1}}$. The beaming factor of each FRB should be $f_{\rm b,FRB} = P f_{\rm b,SGR} < 1/30$. The true energy and true luminosity of FRB 200428 are therefore $E_{\rm FRB}\simeq E_{\rm iso,FRB}f_{\rm b,FRB} < 10^{34}~{\rm erg}$ and $L_{\rm FRB} < L_{\rm iso,FRB}f_{\rm b,FRB} < 3\times10^{36}~{\rm erg~s^{-1}}$, respectively.

\subsection{Constraints on model parameters: Narrow spectra and spectral peak distribution} 

The small $P \leq P_0 = 1/30$ could be also caused by narrow spectra of the putative FRBs, the peak frequency of which has a distribution. Below we constrain the parameter space of this scenario.

We assume that every SGR burst is associated with an FRB-like burst, the peak frequency of which could be outside the FAST band. If the fluence of an FRB reached $f_{\rm FAST}\simeq22~{\rm mJy~ms}$ in the frequency band $\sim(1-1.5)~{\rm GHz}$, it would be detected by FAST. We consider that the FRB peak frequency satisfies a log-normal distribution, that is,
\be
\frac{dN_{\rm FRB}}{d\log\nu_{\rm peak}}=\frac{1}{\sqrt{2\pi}\sigma_{\rm peak}}\exp\left(-\frac{(\log\nu_{\rm peak}-\log\bar\nu_{\rm peak})^2}{2\sigma_{\rm peak}^2}\right),
\ee
where $\nu_{\rm peak}$ and $\bar\nu_{\rm peak}$ are the FRB peak frequency and the mean of its distribution, respectively, and the standard deviation of the distribution is $\sigma_{\rm peak}=0.5$. For each FRB-like event, we assume that the spectrum is a narrow Gaussian, that is,
\be
f_{\nu}=f_{\nu,{\rm peak}}\exp\left(-\frac{(\nu-\nu_{\rm peak})^2}{2\Delta\nu^2}\right),
\ee
where $\Delta\nu$ defines the width of the Gaussian spectrum. The non-detection by FAST then implies
\be
\eta\equiv\frac{f_{\nu,{\rm FRB}}}{f_{\nu,{\rm FAST}}}\simeq\frac{f_{\nu,{\rm peak}}}{f_{\nu,{\rm FAST}}}\lesssim\exp\left(\frac{(\nu-\nu_{\rm peak})^2}{2\Delta\nu^2}\right),
\ee
or
\be
|\nu-\nu_{\rm peak}|\gtrsim(2\ln\eta)^{1/2}\Delta\nu.
\ee
Therefore, the chance probability for non-detection is given by 
\be
P_0\geq P\simeq\int_{\log(\max(1~{\rm GHz}-(2\ln\eta)^{1/2}\Delta\nu,0))}^{\log(1.5~{\rm GHz}+(2\ln\eta)^{1/2}\Delta\nu)}\frac{dN_{\rm FRB}}{d\log\nu_{\rm peak}}d\log\nu_{\rm peak}.
\ee
In \EXTFIG{fig:spectrum}, we plot the relation between $\bar\nu_{\rm peak}$ and $\Delta\nu$ constrained by the observed probability. 
One can see that in order to have $P \leq P_0 = 1/30$, the spectra should be very narrow (for example, $\Delta\nu < 0.1$ GHz) and the distribution peak $\bar\nu_{\rm peak}$ should be far away from the FAST band, for example, either $\bar\nu_{\rm peak}\gtrsim(10-100){\rm GHz}$ or
$\bar\nu_{\rm peak}\lesssim(10-100){\rm MHz}$. In \EXTFIG{fig:spectrum} we show the contours for different probabilities in the $\log \bar\nu_{\rm peak}- \Delta\nu$ plane.

One may use the observation of FRB 200428 to estimate its spectral width. Given that $f_{\nu,{\rm STARE2}}\sim 1.5~{\rm MJy~ms}$ at $\nu_{\rm STARE2}\sim1.4~{\rm GHz}$ and $f_{\nu,{\rm CHIME}}\sim0.22~{\rm MJy~ms}$ at $\nu_{\rm CHIME}\sim0.6~{\rm GHz}$, one can use
\be
\frac{f_{\nu,{\rm STARE2}}}{f_{\nu,{\rm CHIME}}}=\exp\left(\frac{(\nu_{\rm CHIME}-\nu_{\rm STARE2})^2}{2\Delta\nu^2}\right)
\ee
to derive
\be
\Delta\nu=\frac{|\nu_{\rm CHIME}-\nu_{\rm STARE2}|}{[2(\ln f_{\nu,{\rm STARE2}}-\ln f_{\nu,{\rm CHIME}})]^{1/2}}\sim0.4~{\rm GHz}.
\ee

As shown in \EXTFIG{fig:spectrum}, this requires the peak of $\nu_{\rm peak}$ distribution to be far away from the FAST band, that is, $\bar\nu_{\rm peak} \geq 32$ GHz. 

\subsection{Constraints on the fraction of SGRs bursts that produce observable FRBs}

In the following, we show that in order to avoid overproducing the observed FRB rate in the universe, only a small fraction of SGR bursts are allowed to make observable FRBs (not including possible FRB events beaming away from Earth or having narrow spectra outside the gigahertz-level observing window of radio telescopes).

We first estimate the cosmic number density of SGRs as
\begin{eqnarray}
{\cal F}(z)=f_{\rm NS} \cdot f_{\rm SGR}\cdot \int_{\infty}^z\dot{\rho}(z')\cdot \frac{d\tau_{z'}}{dz'}dz',
\end{eqnarray}
where $\dot{\rho}(z)$ is the star formation rate as a function of redshift $z$, $\tau_z$ is the age of the universe at redshift $z$, $f_{\rm NS}$ represents the specific neutron star formation fraction per solar masss ($M_{\odot}$) and $f_{\rm SGR}$ denotes the fraction of neutron stars that can produce SGR bursts (magnetars). We adopt the analytical model of the star formation history derived from the observational data\cite{yuksel08}
\begin{eqnarray}
\dot{\rho}(z)=\dot{\rho}_0\left[(1+z)^{3.4\eta}+\left(\frac{1+z}{5000}\right)^{-0.3\eta}+\left(\frac{1+z}{9}\right)^{-3.5\eta}\right]^{\frac{1}{\eta}},
\end{eqnarray}
where $\dot{\rho}_0=0.02M_{\odot}{\rm yr}^{-1}{\rm Mpc}^{-3}$, and the smoothing parameter $\eta=-10$. The age of the universe for the standard $\Lambda$CDM cosmology is defined as
\begin{eqnarray}
\tau_z=\frac{1}{H_0}\int_z^{\infty}\frac{dz'}{(1+z')[\Omega_M(1+z')^3+\Omega_{\Lambda}]^{1/2}},
\end{eqnarray}
where the Planck cosmological parameters $H_0=67.8{\rm km~s^{-1}Mpc^{-1}}$, $\Omega_m=0.308$ and $\Omega_\Lambda=0.692$ have been used\cite{planck}. Assuming that stars obey the Salpeter initial mass function\cite{salpeter55} $dN/dM\propto M^{-2.35}$ within the mass range $0.1-100 M_{\odot}$ and that massive stars with $8M_{\odot}<M<25M_{\odot}$ would produce neutron stars within a negligible timescale compared with $\tau_z$, we obtain roughly $f_{\rm NS}=0.006 M_{\odot}^{-1}$. In addition, based on Milky Way observations, we can give an order of magnitude estimation for $f_{\rm SGR}\sim N_{\rm SGR,obs}/N_{\rm NS,MW}\sim 10^{-7}$.

To estimate the FRB rate, we further assume that on average SGRs produce hard X-ray bursts with a bursting rate of $\dot{N_b}~\rm day^{-1}$, of which a fraction $f_{\rm SGR-FRB}$ can generate FRBs observable by Earth observers. Consequently, the rate of FRBs from SGRs (SGR-FRBs hereafter) can be estimated as
\begin{eqnarray}
\frac{dN_{\rm FRB}}{dV \cdot dt}={\cal F}(z)\cdot \dot{N_b} \cdot f_{\rm SGR-FRB}.
\end{eqnarray}

For a given detector with the sensitivity threshold $F_{\rm th}$, the threshold luminosity of detectable FRBs at redshift $z$ is $L_{\rm th}=4\pi D_L^2(z)F_{\rm th}$, where $D_L$ is the luminosity distance of the FRB. Here we use $F_{\rm th}\sim \rm 1 mJy$ considering the sensitivity of CHIME\cite{CHIME18}. We assume that the luminosity function of SGR-FRBs is $dN/dL\propto L^{-\alpha}$ within the luminosity range $L_0<L<L_{\rm cut}$, where we take $\alpha=1.8$ and $L_{\rm cut}=10^{44}~{\rm erg~s^{-1}}$ based on the currently known FRB luminosity function\cite{luo20}. We assume $L_0=L_{\rm obs}$, where $L_{\rm obs}=8\times10^{37}~\rm erg~s^{-1}$ is the luminosity of FRB 200428. Finally, we derive the SGR-FRB detection rate 
\begin{eqnarray}
{\cal R_{\rm SGR-FRB}}\sim \int_0^{z_{\rm max}}~{\cal F}(z)\cdot \dot{N_b}\cdot f_{\rm SGR-FRB}\cdot \left(\frac{L_{\rm th}}{L_0}\right)^{-(\alpha -1)}\cdot \frac{dV(z)}{dz}dz,
\end{eqnarray}
where 
\begin{eqnarray}
\frac{dV(z)}{dz}=\frac{c}{H_0}\frac{4\pi D_L^2(z)}{(1+z)^2[\Omega_M(1+z)^3+\Omega_{\Lambda}]^{1/2}},
\end{eqnarray}
and $4\pi D_L^2(z_{\rm max})F_{\rm th}=L_{\rm cut}$. Considering that ${\cal R}_{\rm SGR-FRB}$ should be smaller than the observational rate of FRBs, which is $\sim 10^4~\rm day^{-1}$ all sky\cite{petroff19}, we thus obtain 
\begin{eqnarray}
f_{\rm SGR-FRB}\lesssim 1.7\times 10^{-2}~\left(\frac{{\cal R_{\rm SGR-FRB}}}{10^4~\rm day^{-1}}\right)\left(\frac{f_{\rm SGR}}{10^{-7}}\right)^{-1}\left(\frac{\dot{N_b}}{10^{-4}~\rm day^{-1}}\right)^{-1}.
\end{eqnarray}
This fraction is fully consistent with the fact that at most 1/30 of SGRs can produce observable FRBs.

\subsection{Model constraints from non-detection of FOBs} 

The 17.9-mag upper limit of the optical flux during the prompt phase of FRB 200428 can be used to constrain physical models of FRBs. We consider an extinction correction of 6.2 mag in the direction of SGR 1935+2154. The true upper limit is $\sim$ 11.7 mag.

Following Ref.\cite{yang19}, we consider a putative FOB with peak flux of $F_\nu$ and duration $\tau$. For a telescope with exposure time $T$, the observed effective flux could be estimated as $F_{\nu,{\rm eff}}\sim \min(\tau/T,1)F_\nu$. The magnitude of an optical source is related to its flux through $m=-2.5\log_{10}\left(F_{\nu}/3631~{\rm Jy}\right)$, so
\be
m=
20.8-2.5\log_{10}\left(\frac{\tau_{\rm ms}F_{\nu,{\rm Jy}}}{T_{60}}\right)
\ee
for $\tau\lesssim T$, where $\tau_{\rm ms}$ is the optical pulse duration normalized to milliseconds, $T_{60}$ is the exposure time normalized to $60~{\rm s}$, and $F_{\nu,{\rm Jy}}$ is the peak flux in janskys. 
For a given observed limiting magnitude $m_\ast$, the intrinsic flux limit of an FOB would be 
\be
F_{\nu, {\rm opt}}=
\left(\frac{T_{60}}{\tau_{\rm ms}}\right)10^{(8.32-0.4m_\ast)}~{\rm Jy}
\ee
for $\tau\lesssim T$.
Our BOOTES observation gives an upper limit $m_\ast=11.7$ with $T=60~{\rm s}$ after considering the extinction correction. One then has $F_{\nu,{\rm opt}}\lesssim 4.4~{\rm kJy}$ for $\tau=1~{\rm ms}$. The flux of the FRB 200428 is $F_{\nu,{\rm FRB}}\gtrsim1.5~{\rm MJy}$ for $\tau=1~{\rm ms}$. One therefore has the FOB-to-FRB flux ratio
\be
\zeta\equiv\frac{F_{\nu,{\rm opt}}}{F_{\nu,{\rm FRB}}}\lesssim10^{-3}. 
\ee

This stringent upper limit poses interesting constraints on FOB emission mechanisms\cite{yang19}. For models invoking extension of radio emission in the optical band, the predicted optical flux is lower than this limit. These models are therefore consistent with the upper limit. On the other hand, some models invoking inverse Compton scattering of radio photons to the optical band can be constrained with the current limit. We define the fraction of electrons that can upscatter radio photons to the optical band as $\eta_\gamma \leq 1$. The following constraints can be obtained: (1) For the inverse Compton model within the magnetosphere of a neutron star, $\zeta\lesssim10^{-3}$ leads to the constraint $\eta_\gamma\lesssim3000$. Given that $\eta_\gamma \leq 1$, this scenario is fully consistent with the data. (2) For a beamed radio burst with intrinsic duration $\Delta t$ and opening angle $\theta_j$, and a rotation period of the underlying magnetar of $P\sim3.2~{\rm s}$ sweeping a surrounding nebula, the duration of the FOB due to inverse Compton scattering is much longer, that is, $\tau\sim 1,000~{\rm s}$. The intrinsic optical flux then becomes $F_{\nu,{\rm opt}}\simeq76~{\rm mJy}$, so that 
$\zeta\lesssim10^{-8}$. This gives the constraint 
$\eta_\gamma \lesssim 0.5 (\Delta t/1~{\rm ms})^{-1}$, or $\eta_\gamma \lesssim 1.7\times 10^{-4} (\theta_j/0.1)^{-2}$.
This is the first meaningful constraint on the FOB model parameters.

\clearpage
\clearpage

\section*{Extended Data}

\renewcommand{\baselinestretch}{1.0}
\selectfont

\noindent
\EXTTAB{tab:timeline}: {\bf Information of multi-wavelength campaign.}

\bigskip\noindent
\EXTTAB{tab:fermitab}: {\bf The information of the 29 SGR bursts detected by {\it 
Fermi}/GBM.}

\bigskip\noindent \EXTFIG{fig:limits}: {\bf Flux and fluence upper limits from FAST observation. a,b,} The horizontal axis shows the pulse width, and the vertical axes show the flux ({\bf a}) and fluence ({\bf b}) upper limits.

\bigskip\noindent \EXTFIG{fig:all-radio-SGRburst}: {\bf FRB radio candidates around the epochs of all 29 {\it Fermi}/GBM bursts.} As in \FIG{fig:radio-3SGRburst}, except that the observations are centred around the epochs of different GBM bursts.

\bigskip\noindent \EXTFIG{fig:t90_dis}: {\bf $T_{90}$ and fluence distribution of 29 {\it Fermi}/GBM bursts with best fitting lines.} {\bf a} the duration $T_{90}$ distribution, and {\bf b} the fluence distribution.

\bigskip\noindent \EXTFIG{fig:beaming}: {\bf Jet beaming angle constraints.} {\bf a} the relation between jet beaming angle $\theta_j$ and Lorentz factor 
$\Gamma$ constrained by the observed probability, see Eq.(\ref{thetagamma}).  
{\bf b} the constrained probability ($P$) contours in the $\theta_j-\Gamma$ 
plane. The color scale is in logarithmic scale, i.e., $\log(P)$.

\bigskip \noindent \EXTFIG{fig:spectrum}: {\bf Spectra and spectral peak distribution constraints.} {\bf a} the relation between mean peak frequency $\bar\nu_{\rm peak}$ and spectrum width $\Delta\nu$ constrained by the observed probability. The dashed line corresponds to $\Delta\nu=0.4~{\rm GHz}$.
{\bf b} the constrained probability ($P$) contour in the $\bar\nu_{\rm 
peak}-\Delta\nu$ plane. The color scale is in logarithmic scale, i.e., 
$\log(P)$.

\clearpage
\setcounter{figure}{0}
\setcounter{table}{0}

\captionsetup[table]{name={\bf Extended Data Table}}
\captionsetup[figure]{name={\bf Extended Data Figure}}

\begin{landscape}
\begin{table*}[!htb] \scriptsize
\caption{\bf Information of multi-wavelength campaign \label{tab:timeline}}
\begingroup
\setlength{\tabcolsep}{7pt} 
\renewcommand{\arraystretch}{0.7} 
\begin{tabular}{lcccccccccc}
\hline \hline
band& telescope& location& UTC start & UTC end& Exp & limits \\
& & & & & (s) & \\
\hline
1.25GHz& FAST, Session (1)& Pingtang, China& 2020-04-15T21:54:00& 2020-04-15T23:54:00& 7200 & $<$ 22 mJy ms \\
1.25GHz& FAST, Session (2)& Pingtang, China& 2020-04-26T21:06:55& 2020-04-26T23:06:55& 7200 & $<$ 22 mJy ms \\
1.25GHz& FAST, Session (3)& Pingtang, China& 2020-04-27T23:55:00& 2020-04-28T00:50:37& 3337 & $<$ 22 mJy ms \\
1.25GHz& FAST, Session (4)& Pingtang, China& 2020-04-28T20:35:00& 2020-04-28T23:35:00& 10800 & $<$ 22 mJy ms \\
clear& BOOTES-4 & Lijiang, China& 2020-04-27T18:26:53& 2020-04-27T18:56:53& 
60$\times$30 & $>$ 20.5 mag$^a$ \\
$Z$& BOOTES-2 & Algarrobo Costa, M\'alaga, Spain& 2020-04-28T00:44:03& 2020-04-28T00:48:14& 5$\times$50 & $>$ 15.9 mag \\
$Z$ & BOOTES-3 & NIWA Lauder, Otago, New Zealand& 2020-04-28T14:34:24& 2020-04-28T14:35:24& 60 & $>$ 17.9 mag \\
clear & BOOTES-2 & Algarrobo Costa, M\'alaga, Spain & 2020-04-28T22:57:44& 2020-04-28T23:27:44& 60$\times$30 & $>$ 20.4 mag \\
$R$ & LCOGT & Fort Davis, US& 2020-04-30T07:21:49& 2020-04-30T08:07:08& 8$\times$300 & $>$ 21.1 mag \\
$1-250$ keV& Insight-HXMT& ...& 2020-04-26T20:14:50& 2020-04-27T00:40:56& 3$\times$3675 & $<({\rm LE:}3.8, {\rm ME:}4.9, {\rm HE:}6.4)\times 10^{-9}~{\rm erg~cm^{-2}}$ \\
$1-250$ keV& Insight-HXMT& ...& 2020-04-28T07:27:03& 2020-04-29T11:30:02& 18$\times$3738 &$<({\rm LE:}3.8, {\rm ME:}4.9, {\rm HE:}6.4)\times 10^{-9}~{\rm erg~cm^{-2}}$ \\
\hline \end{tabular}
\endgroup
\end{table*}
$^a$ All optical magnitude upper limits are subject to extinction correction, which is 6.2 mag in $Z$ band for SGR J1935+2154.
\end{landscape}

\begin{longtable}{l c c c c}%
\caption{\bf The information of the 29 SGR bursts detected by {\it Fermi}/GBM 
\label{tab:fermitab}}
\\
\hline%
\hline%
ID&Burst time&Flux (erg cm$^{-2}$ s$^{-1}$)&Fluence (erg cm$^{-2}$)\\%
\hline%
\endhead%
\hline%
\endfoot%
\hline%
\endlastfoot%
1&00:19:44.192&1.22$_{-0.16}^{+0.18} \times 10^{-6}$&7.93$_{-1.05}^{+1.20} \times 10^{-8}$\\%
2&00:23:04.728&3.26$_{-0.67}^{+0.73} \times 10^{-7}$&7.10$_{-1.47}^{+1.58} \times 10^{-8}$\\%
3&00:24:30.296&2.37$_{-0.05}^{+0.05} \times 10^{-5}$&3.01$_{-0.07}^{+0.06} \times 10^{-6}$\\%
4&00:25:43.945&1.99$_{-0.62}^{+0.70} \times 10^{-7}$&5.26$_{-1.64}^{+1.84} \times 10^{-8}$\\%
5&00:37:36.153&2.73$_{-0.55}^{+0.58} \times 10^{-7}$&6.78$_{-1.35}^{+1.43} \times 10^{-8}$\\%
6&00:39:39.513&8.96$_{-1.04}^{+1.09} \times 10^{-7}$&1.89$_{-0.22}^{+0.23} \times 10^{-7}$\\%
7&00:40:33.072&1.20$_{-0.11}^{+0.11} \times 10^{-6}$&3.57$_{-0.33}^{+0.32} \times 10^{-7}$\\%
8&00:41:32.136&4.69$_{-0.17}^{+0.16} \times 10^{-6}$&1.15$_{-0.04}^{+0.04} \times 10^{-6}$\\%
9&00:43:25.169&2.23$_{-0.13}^{+0.14} \times 10^{-6}$&5.51$_{-0.33}^{+0.35} \times 10^{-7}$\\%
10&00:44:08.202&3.93$_{-0.07}^{+0.08} \times 10^{-5}$&6.68$_{-0.13}^{+0.13} \times 10^{-6}$\\%
11&00:45:31.097&8.44$_{-1.16}^{+1.32} \times 10^{-7}$&7.43$_{-1.02}^{+1.16} \times 10^{-8}$\\%
12&00:46:00.009&7.83$_{-0.67}^{+0.66} \times 10^{-7}$&4.52$_{-0.39}^{+0.38} \times 10^{-7}$\\%
13&00:46:06.408&4.11$_{-0.61}^{+0.64} \times 10^{-7}$&7.89$_{-1.17}^{+1.23} \times 10^{-8}$\\%
14&00:46:20.176&2.32$_{-0.05}^{+0.06} \times 10^{-5}$&4.18$_{-0.09}^{+0.10} \times 10^{-6}$\\%
15&00:46:23.504&3.17$_{-0.43}^{+0.46} \times 10^{-7}$&2.32$_{-0.31}^{+0.33} \times 10^{-7}$\\%
16&00:46:43.208&9.81$_{-0.69}^{+0.75} \times 10^{-7}$&3.21$_{-0.23}^{+0.24} \times 10^{-7}$\\%
17&00:47:24.961&1.66$_{-0.34}^{+0.43} \times 10^{-7}$&6.23$_{-1.29}^{+1.61} \times 10^{-8}$\\%
18&00:47:57.528&1.16$_{-0.11}^{+0.12} \times 10^{-6}$&1.08$_{-0.10}^{+0.11} \times 10^{-7}$\\%
19&00:48:44.824&3.96$_{-0.42}^{+0.46} \times 10^{-7}$&1.38$_{-0.15}^{+0.16} \times 10^{-7}$\\%
20&00:48:49.272&3.05$_{-0.16}^{+0.17} \times 10^{-6}$&7.32$_{-0.38}^{+0.40} \times 10^{-7}$\\%
21&00:49:00.273&7.80$_{-1.03}^{+1.14} \times 10^{-7}$&8.11$_{-1.07}^{+1.18} \times 10^{-8}$\\%
22&00:49:01.121&8.36$_{-0.92}^{+0.96} \times 10^{-7}$&1.32$_{-0.14}^{+0.15} \times 10^{-7}$\\%
23&00:49:06.472&9.66$_{-3.73}^{+4.00} \times 10^{-8}$&6.98$_{-2.69}^{+2.89} \times 10^{-8}$\\%
24&00:49:16.592&1.78$_{-0.15}^{+0.16} \times 10^{-7}$&4.17$_{-1.24}^{+1.28} \times 10^{-8}$\\%
25&00:49:22.392&7.72$_{-1.05}^{+1.10} \times 10^{-7}$&4.55$_{-0.62}^{+0.65} \times 10^{-8}$\\%
26&00:49:27.280&3.58$_{-1.08}^{+1.32} \times 10^{-7}$&2.11$_{-0.64}^{+0.78} \times 10^{-8}$\\%
27&00:49:46.680&3.87$_{-0.36}^{+0.39} \times 10^{-7}$&2.63$_{-0.24}^{+0.26} \times 10^{-7}$\\%
28&00:50:01.248&7.83$_{-0.61}^{+0.66} \times 10^{-7}$&3.13$_{-0.25}^{+0.26} \times 10^{-7}$\\%
29&00:50:21.969&1.32$_{-0.46}^{+0.55} \times 10^{-7}$&1.85$_{-0.65}^{+0.76} \times 10^{-8}$\\%
\end{longtable}

\clearpage
\begin{figure}
 \centering
 \includegraphics[width=4.5in]{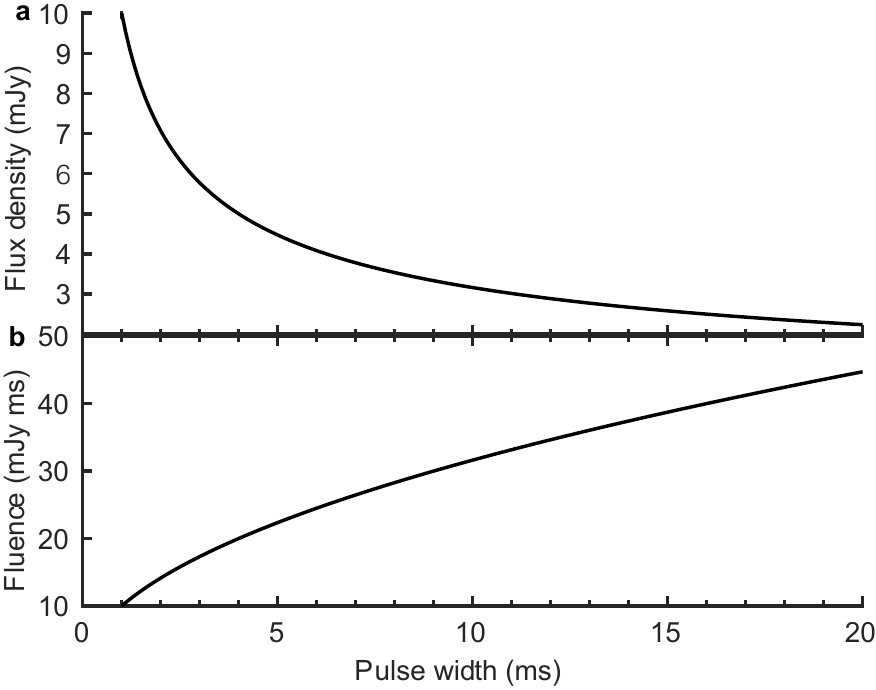}
 \caption{{\bf Flux and fluence upper limits from FAST observation. a,b,} The horizontal axis shows the pulse width, and the vertical axes show the flux ({\bf a}) and fluence ({\bf b}) upper limits. \label{fig:limits}}
\end{figure}

\begin{figure}
 \centering
 \includegraphics[width=7in]{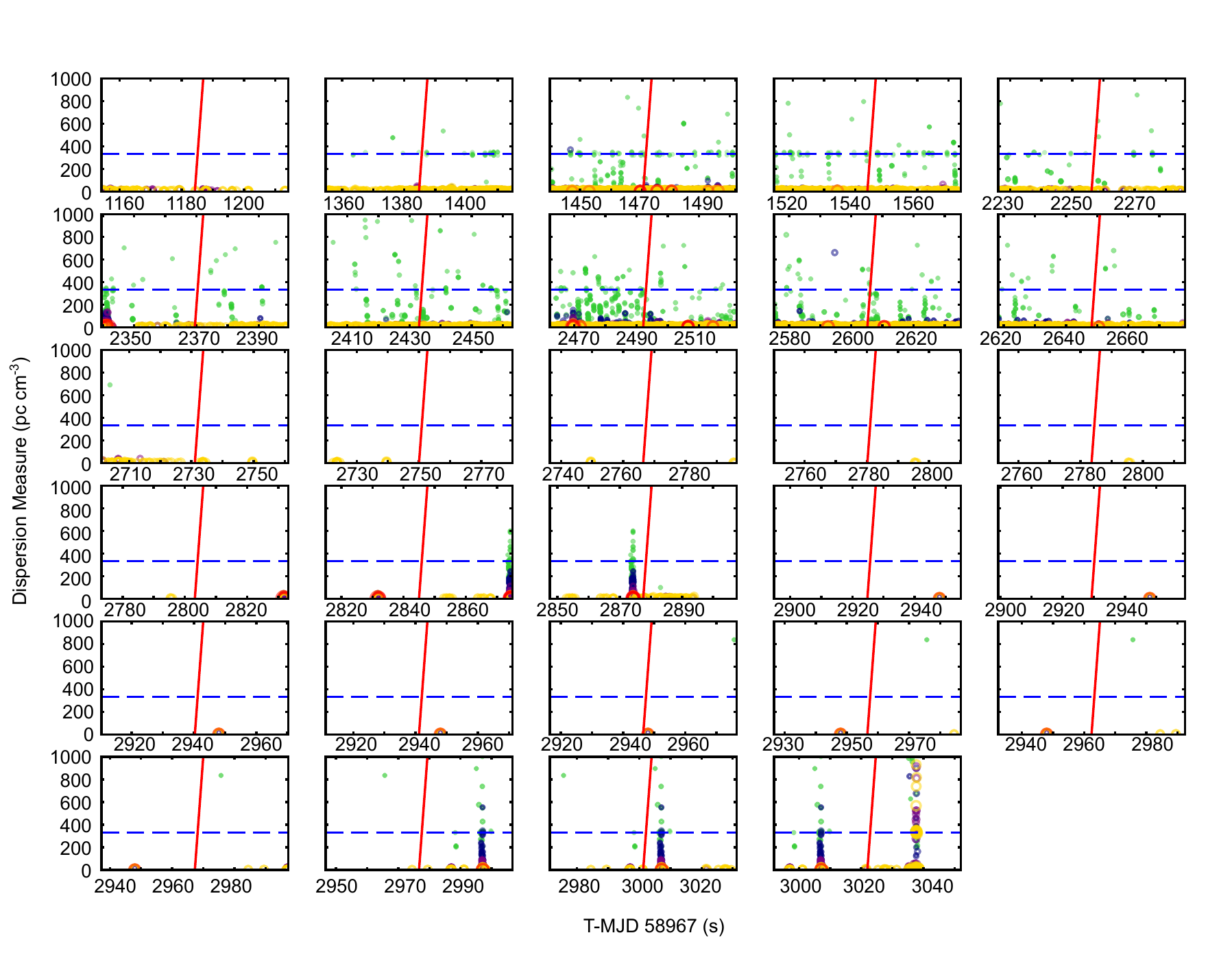}
 \caption{{\bf FRB radio candidates around the epochs of all 29 {\it Fermi}/GBM bursts.} As in \FIG{fig:radio-3SGRburst}, except that the observations are centred around the epochs of different GBM bursts.  
 \label{fig:all-radio-SGRburst}}
\end{figure}

\begin{figure}
 \centering
 \includegraphics[width=7.0 in]{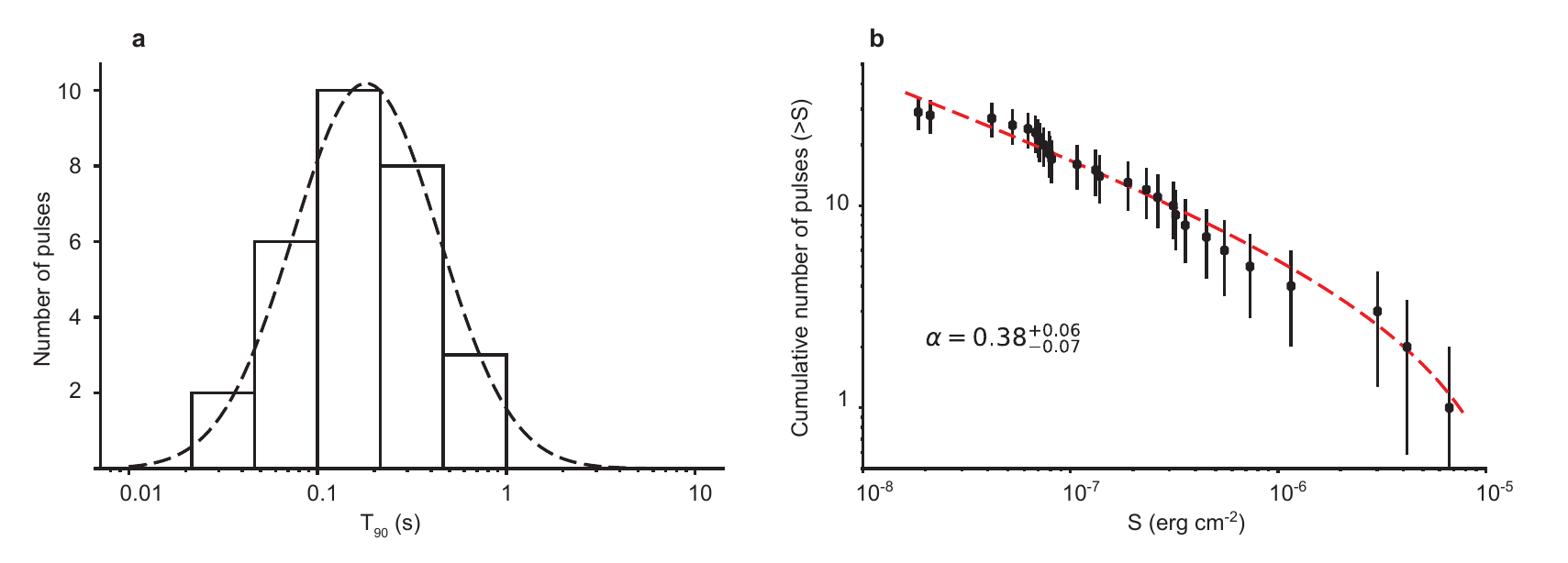}
 \caption{{\bf $T_{90}$ and fluence distribution of 29 {\it Fermi}/GBM bursts with best fitting lines.} {\bf a} Distribution of the duration $T_{90}$. {\bf b} Fluence distribution. The errors are defined as the square root of the burst number in each fluence bin.
 \label{fig:t90_dis}}
\end{figure}

\begin{figure}
\centering
\includegraphics[width=7.0 in]{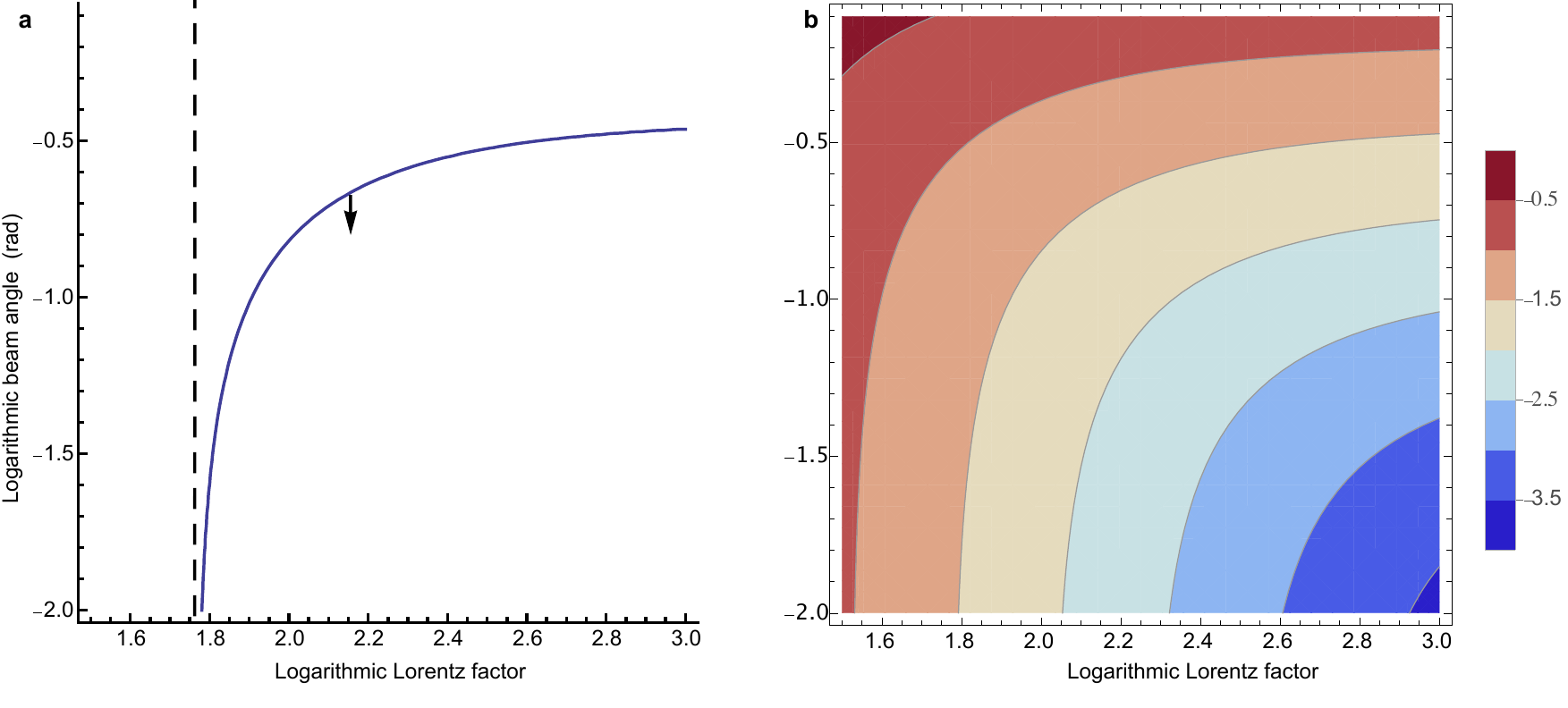}
\caption{{\bf Jet beaming angle constraints.} {\bf a} Relationship between jet beaming angle $\theta_j$ and Lorentz factor $\Gamma$ constrained by the 
observed probability, see Eq.(\ref{thetagamma}). {\bf b}  Constrained probability ($P$) contours in the $\theta_j-\Gamma$ plane. The colour scale is  
logarithmic, that is, $\log(P)$.\label{fig:beaming}} \end{figure}

\begin{figure}
\centering
\includegraphics[width=7.0 in]{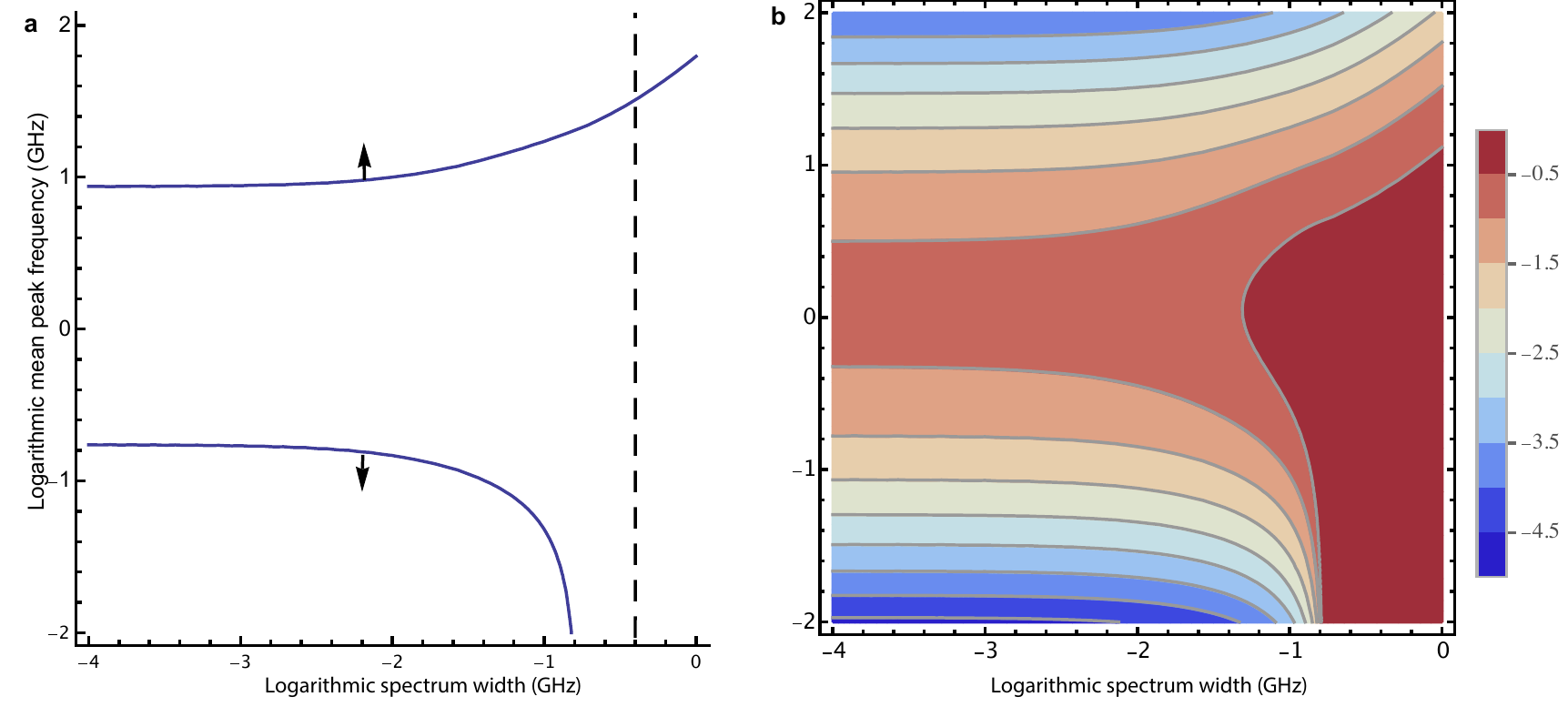}
\caption{{\bf Spectra and spectral peak distribution constraints.} {\bf a} Relationship between mean peak frequency $\bar\nu_{\rm peak}$ and spectrum width $\Delta\nu$ constrained by the observed probability. The dashed line corresponds to $\Delta\nu=0.4~{\rm GHz}$. {\bf b} Constrained probability ($P$) contours in the $\bar\nu_{\rm 
peak}-\Delta\nu$ plane. The colour scale is in logarithmic scale, that is, $\log(P)$.\label{fig:spectrum}}
\end{figure}

\end{document}